# ExoCat-1: The Nearby Stellar Systems Catalog for Exoplanet Imaging Missions

Margaret C. Turnbull
SETI Institute, Carl Sagan Center for the Study of Life in the Universe
turnbull.maggie@gmail.com

## ABSTRACT

We present the first version of a Nearby Stellar Systems Catalog for Exoplanet Imaging Missions (dubbed by the direct imaging community as "ExoCat") for use in exoplanet direct imaging mission planning. This version, ExoCat-1, includes 2347 stars taken from the *Hipparcos* Catalogue with measured parallaxes of $\pi > 33.33$ mas (corresponding to a distance of 30 pc). This sample is nearly complete down to V=8, corresponding to stars brighter than $\sim 0.5$ $L_{\odot}$ (corresponding to late G-/early K-type dwarf stars at the 30 pc distance limit). For each star we provide astrometry (including Equatorial and Galactic coordinates, parallax, and proper motions), Johnson *B* and *V* magnitudes (converted from *Hipparcos* or Tycho data or taken from the literature), and $K_s$-band magnitudes from *2MASS* (for fainter stars) or *K*-band magnitudes taken from the literature and converted to *2MASS* $K_s$ magnitudes (for bright stars). Using these data we estimate stellar luminosity, effective temperature, stellar radius (in solar and angular units), Earth-equivalent insolation distances (in AU and in angular units), and fractional planet brightness for an exo-Earth at the Earth-equivalent insolation distance. We provide published spectral types and simple labels on stellar type for quick assessment of design reference mission-selected target lists (e.g., "O", "A", "B", "F", "G", "K", "M", "WD," "SUB," and "GIANT"). The number of known exoplanet companions is indicated for each star, and for bright stars (V<7) we provide separations and delta-magnitudes for the brightest stellar companion within 10 arcseconds. Other important stellar data such as log(g), metallicity, chromospheric activity level, and age estimates are provided where readily available. ExoCat-1 can be found through the Exoplanets Exploration Program (ExEP) website.

## 1. Motivation for ExoCat

In twenty years, every planet detection technique employed has had a surprising discovery to its credit, pointing to a previously unimagined diversity in both planet hosts (including K through F main sequence stars, eclipsing binaries, wide binaries, giants, white dwarfs, and pulsars) and in planetary characteristics (including super-Earths, mini-Neptunes, hot giants, and retrograde planets; Howard 2013). These findings have bolstered the case for a dedicated direct imaging mission to find and spectrally characterize planets orbiting the nearest stars, an important step in understanding their true nature (Laughlin & Lissauer 2015; Burrows 2014).

In 2010, the "New Worlds, New Horizons" Astrophysics Decadal Survey identified technology development to study nearby Earth-like exoplanets as the highest medium-scale programmatic priority for this decade (NRC 2010). Responding to this, in 2013 NASA's Exoplanet Exploration Program (ExEP) commissioned three Science and Technology Definition Teams to explore two reduced-scale exoplanet imaging concepts (an internal coronagraph and an external occulter or "starshade") and a coronagraph for the upgraded WFIRST AFTA mission. The final releases of all three studies were published in early 2015 (Stapelfeldt et al, 2015; Seager et al. 2015; Spergel et al. 2015).

Defining performance requirements and assessing science yields for all such mission concepts requires detailed information about their target stars. For example, in the case of "discovery" targets wherein no planets are currently known, habitable zone locations drive inner working angle and starlight suppression requirements, spectral energy distributions indicate how planets will appear in reflected starlight as compared to faint background sources, stellar companion separations and relative brightnesses are needed to estimate stray light contamination in the field of view, galactic coordinates predict the density and type of background sources likely to be found at planet-like brightnesses (V ~ 25-30), stellar age may predict debris and dust levels, and proper motions indicate the minimum time between observations in order to establish common motion between sources as well as to estimate the probability that any detected planets will still be visible upon a second visit. Meanwhile, stellar characteristics such as abundances, activity, age, and space motion are of scientific interest in exploring how such parameters might be correlated with the prevalence and diversity of planets.

In an effort to create a single, user-friendly database of reliable information for the stars within 30 pc, we have developed the Nearby Stellar Systems Catalog for Exoplanet Imaging Missions (dubbed "ExoCat" by its users). This first version, ExoCat-1, includes 2347 entries for *Hipparcos* stars located within 30 pc of the Sun. We provide astrometry (including Equatorial and Galactic coordinates, parallax, and proper motions), Johnson *B* and *V* magnitudes (converted from *Hipparcos* or *Tycho* data or taken from the literature), and $K_s$-band magnitudes from *2MASS* (for fainter stars) or *K*-band magnitudes taken from the literature and converted to *2MASS $K_s$* magnitudes (for bright stars). Using these data we estimate stellar luminosity, effective temperature, stellar radius (in solar and angular units), stellar mass, habitable zone locations (in AU and in angular units), and fractional planet brightness for an exo-Earth at the Earth-equivalent insolation distance. We provide published spectral types and simple labels on stellar type for quick assessment of design reference mission (DRM) selected target lists (e.g., "O", "A", "B", "F", "G", "K", "M", "WD" and "GIANT"). The presence of known exoplanet companions is indicated, and for bright stars (V<7) we provide separations and delta-magnitudes for the brightest stellar companion within 10 arcseconds, taken from the Washington Double Star catalog. Other important stellar data such as log(g), metallicity, chromospheric activity level, and age estimates are provided where readily available. ExoCat-1 can be downloaded as an Excel spreadsheet from the Exoplanets Exploration Program (ExEP) website (http://nexsci.caltech.edu/missions/EXEP/EXEPstarlist.html). The Catalog is expected to be expanded in the near future to include later spectral types, additional photometric measurements, and additional companion data.

In creating ExoCat-1, we made extensive use of the Vizier online service, which allows users to query thousands of datasets in the astronomical literature, including the use of constraints on individual parameters and matching to user-created input lists. In Section 2, we describe the stars and compilation of astrometry and photometry, spectral types, and other observed parameters. In Section 3, we describe the derivation of other quantities of interest such as stellar luminosities and radii, Earth-equivalent insolation distance (EEID), and estimates of the magnitudes and fractional planet brightness of hypothetical Earth analogs. In Section 4 we describe how ExoCat-1 will be maintained and future improvements incorporated.

## 2. The Stars of ExoCat

Our primary goal in building ExoCat is straightforward: create a compilation of photometric and astrophysical data for all stars within 30 pc. These stars have been identified based on parallax measurements, which is by far the most accurate method of determining a star's distance (Henry et al. 2006). A star's distance, in turn, is a crucial input in estimating stellar luminosity, angular radius, habitable zone location, habitable zone angular size, and predicted fractional planet brightnesses – all of which come into play when determining the best targets for direct imaging of exoplanets.

For the 30 pc sample in the *Hipparcos* Catalog, ExoCat-1 contains HIP numbers and identifiers from the Annie Jump Cannon's Henry Draper Catalog ("HD;" Cannon & Pickering 1922), the Nearby Stars catalog ("GL" for numbers < 1000 and "GJ" for numbers > 1000, "NN" for numbers 3000-5000; Gliese & Jahress 1995), the Revised Luyten Half Second Catalog (LHS; Luyten 1976; Bakos et al. 2002), the Luyten two-tenths catalog (LTT; Luyten 1979), the Washington Double Star Catalog (WDS; Mason et al. 2001 and subsequent updates) and common names (for the brightest stars). We also include a smattering of other identifiers taken from the *Hipparcos* Catalog. HIP numbers do not denote "A," "B," etc, for components of multiple systems, but this lettering was preserved in the other identifiers. Given that many users in the space instrumentation industry do not intuitively use the more traditional identifiers preferred by astronomers (especially HD and GL/GJ), we opted to avoid confusing the situation further and have not included an additional ExoCat designation. Equatorial coordinates and proper motions are reproduced from Hipparcos, and we use these RA and Dec to calculate Galactic coordinates. The following describes the parallax and photometry included in ExoCat-1, with a discussion of uncertainties.

### 2.1 Parallax Data

The *Hipparcos* Catalogue (ESA 1997) is the largest single source of high precision astrometric measurements, and the sample is nearly complete out to 30 pc for stars of spectral type ~K5V and earlier.  For this reason, *Hipparcos* is the most logical starting point for a compilation of stellar data for bright stars, and will remain so until the production of the *Gaia* mission's much anticipated final catalog of high precision astrometric, photometric, and kinematic data for 1 billion stars down to ~20th magnitude (see Jordi 2014 for a review of the *Gaia*'s post-launch performance).

This initial version of ExoCat, which we refer to as ExoCat-1, contains only stars that have *Hipparcos* parallax measurements (with π>0.03333").  This compilation has been in casual circulation for several years and used for a variety of studies up to this point (e.g., Seager et al. 2015, Léger et al. 2015, Brown 2015, Turnbull et al. 2012, Brown 2006).  For mission studies focusing on the nearest Sun-like systems, ExoCat-1 is expected to be useful for early estimates of science yields.

However, using a list based on the *Hipparcos* Catalogue as an input list for missions to directly image exoplanets has important limitations.  *Hipparcos* was not an all sky survey, but rather relied on a predetermined Input Catalog (HIC) of ~118,000 bright stars constructed from a magnitude limited survey of ~58,000 stars plus an additional list of high priority targets proposed by the scientific community (Turon 1992).  Thus, while the *Hipparcos* limiting magnitude is V ~ 12, the catalog is *complete* only to V ~ 7.9 + 1.1sin|$b$| for spectral types earlier than G5, and V<7.3 + 1.1sin|$b$| for spectral types later than G5, where $b$ is Galactic latitude (ESA 1997).

As a result, the completeness of the 30 pc sample falls off dramatically for fainter stars, and as a result leaves out perhaps half of the members of the Solar Neighborhood: the ubiquitous M dwarfs (Winters et al. 2015; Dittman et al. 2015).  Approximately half of the nearest 100 stellar systems components are missing from *Hipparcos* due to this magnitude limit (RECONS 2012).  There are also a few bright, potentially important target stars that were overlooked in compiling the *Hipparcos* Input Catalog (e.g. the 6th magnitude K2V star GL 216 B, a wide companion to the 3rd magnitude F6V star GL 216 A, HIP 27072).  Multiple systems were especially challenging for the *Hipparcos* mission, as many systems were unresolved prior to flight and oftentimes the resulting Catalogue entries refer to more than one star or leave secondary components unaccounted for (Lindegren et al. 1997; Dommanget & Nys 2000).  Finally, not *all* of the astrometry contained in the *Hipparcos* Catalogue is reliable, especially when two or more bright stars share the field (van Leeuwen 2007).  For these reasons, design reference mission (DRM) studies using the *Hipparcos* Catalogue as an input list should be regarded as preliminary.  High priority targets should be subjected to careful individual scrutiny before proposing observational programs for upcoming exoplanet imaging missions, such as for the WFIRST coronagraph.

Using Vizier to query van Leeuwan's (2007) new reduction of *Hipparcos* astrometric data returns 2348 stars with π ≥ 33.33 milliarcseconds, from which we removed one known spurious entry, HIP 114110, for a total of 2347 stars in the ExoCat-1 sample. Typical parallax uncertainties for these stars are ~ 2 mas, with fractional uncertainties $\sigma_\pi/\pi$ <1% for stars brighter than V~6. For fainter stars, and for stars with nearby bright optical or physical companions, parallax uncertainty can be significantly higher. Parallax uncertainty is typically the greatest source of error in calculating stellar luminosities. Figure 1 shows the distribution of fractional parallax uncertainties (solid line) and cumulative distribution of parallax uncertainties in percent (dashed line).

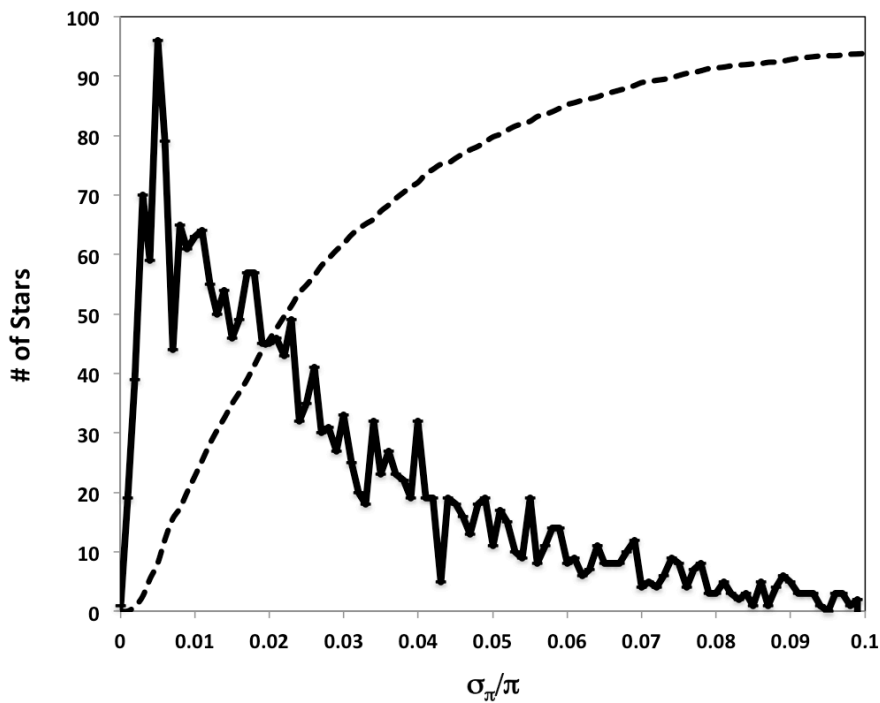

Figure 1. Fractional Parallax Uncertainties for stars included in Exocat-1 (solid line) and cumulative percent (dashed line). Approximately half the stars in ExoCat-1 have parallax uncertainties of ~2% or better.

## 2.2 Photometry

Our approach to finding stellar luminosities is to use $B$, $V$, and $K_s$-band photometry plus fits to the tables of intrinsic colors of O9-M9 dwarfs provided by Pecaut & Mamajek (2013) to estimate bolometric corrections to $V$-band data. In the following subsections, we describe the $BVK_s$ photometry included in ExoCat-1, its use in calculating stellar luminosities, and the effects of parallax and photometric uncertainty on the HR diagram.

The $V$ and $B–V$ photometry provided in the *Hipparcos* Catalogue comes either from ground-based measurements made prior to launch, or from the on-board *Hipparcos*-Tycho $H_pB_TV_T$ System. $H_p$ magnitudes (column H44 in *Hipparcos*) correspond to a wide, unfiltered, visible light bandpass, and $V_T$ and $B_T$ (columns H32 and H34) correspond roughly to Johnson $V$ and $B$. For stars brighter than ~8.5, the $H_pB_TV_T$ system measurements typically have precisions of a few tenths of a millimagnitude (Bessell 2005), but the Tycho magnitudes given in *Hipparcos* have subsequently been revised and are published in the Tycho-2 Catalog with improved precision (Høg et al. 2000). The *Hipparcos* Catalogue published equations for transforming $H_p$, $V_T$, and $B_T$ magnitudes to the Johnson system based on pre-flight calibrations, but these relations were also revised by Bessell (2000) to correct for systematic differences possibly arising from changes in the detector response at shorter wavelengths resulting from repeated passage through the Van Allen radiation belts (Bessell 2005).

Our goal was to include the most precise *B* and *V* magnitudes possible, whether they originate from the space- or ground-based measurements. The V-band magnitudes quoted by *Hipparcos* for the 2347 stars in the ExoCat-1 sample includes 1334 pre-flight ground-based measurements (flagged "G" in *Hipparcos*), 1012 measurements converted from $H_p$ ("H"), and 1 measurement converted from $V_T$ ("T"). Meanwhile, the *B–V* colors quoted for these stars include 1977 ground-based measurements ("G"), and 311 measurements converted from $B_T – V_T$ measurements ("T").

Rather than simply adopting these values, we chose to recalibrate the satellite data to the Johnson system using polynomial fits to Table 2 in Bessell (2000) and the updated Tycho-2 photometry, unless the ground-based measurements had lower quoted uncertainties. $H_p$ satellite magnitudes are given in *Hipparcos* for all 2047 HIP30 stars, $V_T$ magnitudes for 1772 stars, and $B_T$ magnitudes for 1766 stars. *V–I* colors, which can be used in the transformation from $H_p$ to *V* when no *B*-band data are available, were also given for 2288 stars (although the given *V–I* magnitudes are sometimes problematic for redder stars; see Platais et al. 2003 and Vizier for a useful list of known errors in the *Hipparcos* Catalogue; http://vizier.cfa.harvard.edu/viz-bin/getCatFile?-plus=-%2b&I/239/errata.htx).

Additionally, an important recent source of high precision ground-based Johnson photometry for fainter stars (V > 6.5 mag) comes from the southern ($\delta < 26°$) survey at SAAO by Koen et al. (2010), which we generally preferred to use over converting to Johnson *B* and *V* from satellite magnitudes. To determine whether to use ground-based or satellite data for the *V* and *B–V* magnitudes in ExoCat, we simply added the quoted $V_T$ and $B_T$ uncertainties in quadrature. If this uncertainty was smaller than the uncertainty quoted for the given ground-based *B–V* data, we opted to use the *Tycho*-2 photometry.

### 2.2.1 Johnson *V*-band Photometry.

Of the 2347 stars in HIP30, we found 526 stars in Koen et al. (2010; V_src = "Koen" in ExoCat), with quoted mean uncertainties as good as or better than the Tycho photometry or older ground-based measurements ($\sigma_{avg}$ = 0.012 mag in *V* band). Additionally, there are 885 stars with pre-*Hipparcos* ground-based *V*-band data (V_src = "G" in ExoCat) with quoted uncertainties as good as or better than the corresponding satellite data. These quoted *V* magnitudes and uncertainties are used directly in ExoCat. If only *B–V* uncertainties were available, we divided the *B–V* uncertainty by $\sqrt{2}$ to estimate the uncertainty in *V* ("e_V" in ExoCat).

For the bulk (899) of the remaining stars, we converted $V_T$ to $V$ magnitudes using *Tycho*-2 photometry and Equation (1), taken from Mamajek et al. (2006; V_src = "VT" in ExoCat-1, or "VT-G" to note that ground-based data were originally reported in *Hipparcos*). The residuals of this fit to Bessell's (2005) Table 2 are less than 0.6 millimagnitudes for the range $-0.25 < B_T - V_T < 2.0$. We estimated the uncertainty in $V$ for these stars by adding the given $V_T$ and $B_T$ standard errors in quadrature.

For 20 stars lacking Tycho-2 $B_T$ measurements, we converted $H_p$ to $V$ using *Hipparcos*' V-I values (where V-I comes either from pre-flight ground-based measurements in column H40 or was estimated from $H_p$ and $V_T$ for the purpose of data reductions in column H75), as in Equation (2). The residuals of this fit to Bessell's table are less than 0.008 magnitudes within the range $-0.25 \leq B_T - V_T \leq 2.5$. In cases where the *V-I* photometry used in Equation (2) reflects a ground-based measurement (V_src = "Hp-VI"), the uncertainty in $V$ quoted in ExoCat-1 is the uncertainties for $H_p$ and *V-I* added in quadrature. For cases where the *V-I* photometry is merely itself an estimate used for *Hipparcos* data reductions (V_src = "Hp-VI-red"), we quote the *V*-band uncertainty as ">0.1."

For 6 additional stars lacking Tycho-2 $B_T$ measurements and having V-I color redder than 2.5, we simply kept the V-band magnitude and uncertainty originally given in Hipparcos. Finally, for 11 stars with no Tycho-2 $B_T$ magnitude available (or very high $\sigma_{BT}$) and no other $B-V$ data available, we deemed the conversions from $H_p$ to $V$ given in Hipparcos to be unreliable. For these stars we provide $V$-band magnitudes found in SIMBAD and either quote the uncertainty given there or list e_V = ">0.1."

In summary, the equations used to convert $V_T$ and $H_p$ magnitudes to Johnson $V$ magnitudes, and to quote uncertainties in ExoCat-1, are as follows:

<u>$H_p$ to Johnson $V$ (using $B_T - V_T$):</u>

$$V = V_T \quad - 0.00097 \hspace{4cm} (1)$$

$$\quad\quad - 0.13342 \, (B_T - V_T)$$

$$\quad\quad + 0.05486 \, (B_T - V_T)^2$$

$$\quad\quad + 0.01998 \, (B_T - V_T)^3$$

where uncertainty in $V$ is given as e_V = $\sqrt{(\sigma_{BT}^2 + \sigma_{VT}^2)}$ mag, and V_src = "VT."

<u>$H_p$ to Johnson $V$ (using $V - I$):</u>

$$V = H_p \quad - 0.0127$$

$$\quad\quad - 0.2788 \, (V - I)$$

$$\quad\quad + 0.4037 \, (V - I)^2$$

$$- 0.6809 \, (V - I)^3 \qquad\qquad (2)$$

$$+ \; 0.6187 \, (V - I)^4$$

$$- 0.2420 \, (V - I)^5$$

$$+ \; 0.0330 \, (V - I)^6$$

where uncertainty in $V$ is given as: e_V = $\sqrt{(\sigma_{Hp}^2 + \sigma_{V-I}^2)}$ mag (if V_src = "Hp-VI"), and uncertainty in $V$ is given as e_V = ">0.1" (if V_src = "Hp-VI-red").

### 2.2.2 Johnson *B−V* Colors

The uncertainties in *Tycho*-2 $B_T$ magnitudes are generally larger than for $V_T$ (and indeed the uncertainties in both bands are generally larger than the uncertainties originally published *Tycho*, which were underestimated; Høg et al. 2000). As a result, the majority of stars in ExoCat-1 have *B−V* colors taken from higher quality ground-based data. We found 526 stars with *B−V* data in Koen et al. (2010), who quote a mean uncertainty of $\sigma_{B-V}$ = 0.0084 magnitudes (B-V_src = "Koen" in ExoCat-1 and e_B-V = 0.0084 mag). For 1408 stars, ExoCat-1 reproduces the B-V measurements and associated uncertainties given in *Hipparcos* (B-V_src = "G").

For 389 stars where the *Tycho*-2 photometry appears to be more precise than the ground-based data (or where ground-based data were not available), we used the Mamajek et al. (2006) equations to convert from $B_T − V_T$ to $B−V$. These relations are reproduced below. Equation (3) was used for stars with $-0.25 < B_T − V_T < 0.4$ (noted in ExoCat-1 as B-V_src = "BVT-Blue," or "BVT-G-Blue" to indicate that ground-based data are also available), and Equation (4) was used for stars with $0.4 < B_T − V_T < 2.0$ (noted in ExoCat-1 as B-V_src = "BVT-Red," or "BVT-G-Red" to indicate that ground-based data are also available). The residuals of these fits to Bessell's (2000) tables are less than a few millimagnitudes for these color ranges and uncertainties in $B−V$ are simply estimated as e_B-V = $\sqrt{(\sigma_{BT}^2 + \sigma_{VT}^2)}$.

The above accounts for nearly all of the stars in ExoCat-1. For the remaining 24 stars, no trustworthy photometry is given in Hipparcos. These stars are all relatively faint (V ∼ 11-13) and many have close visual companions that compliate the photometry. For half of these (12 stars), we did find *B−V* data in SIMBAD (B-V_src = "SIMBAD;), and we either quote the uncertainty given there or list e_V = ">0.1." For the final 12 stars, no *B−V* data were found.

The relations used to convert from $B_T − V_T$ to Johnson *B−V* come from Mamajek et al. (2006) and are as follows:

For $-0.25 < B_T − V_T < 0.4$:

$$B−V \; = \quad B_T − V_T$$

$$- 0.006$$

$$– 0.1069 \; (B_T – V_T) \tag{3}$$

$$+ 0.1459 \; (B_T – V_T)^2 \; ,$$

and e_B–V = $\sqrt{(\sigma_{BT}^2 + \sigma_{VT}^2)}$ mag.

For $0.4 < B_T – V_T < 2.0$:

$$B–V \quad = \quad B_T – V_T$$

$$– 0.007813 \; (B_T – V_T)$$

$$– 0.1489 \; (B_T – V_T)^2 \tag{4}$$

$$+ 0.03384 \; (B_T – V_T)^3 \; ,$$

and e_B–V = $\sqrt{(\sigma_{BT}^2 + \sigma_{VT}^2)}$ mag.

Figures 2 and 3 show the full range of $V$ and $B–V$ for stars in ExoCat-1, with associated uncertainties.

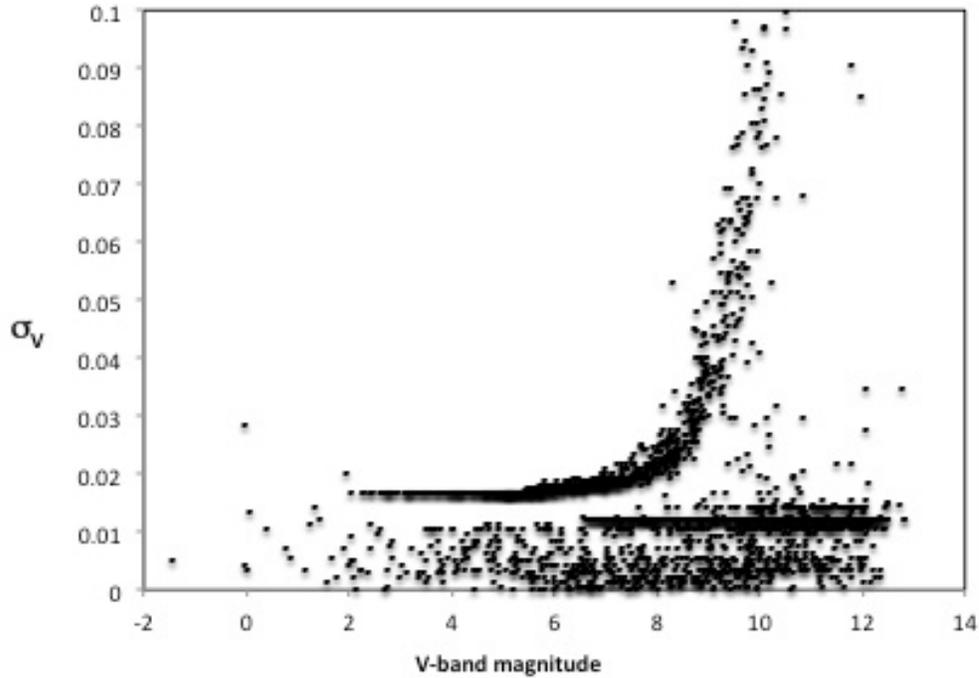

Figure 2. Range of V-band magnitudes in ExoCat-1, and associated uncertainties for the various sources of photometry. Only ~160 stars have uncertainties in $B–V$ higher than ~3%, and 88 stars higher than 0.1 mag (including the 12 stars with no B-V data available).

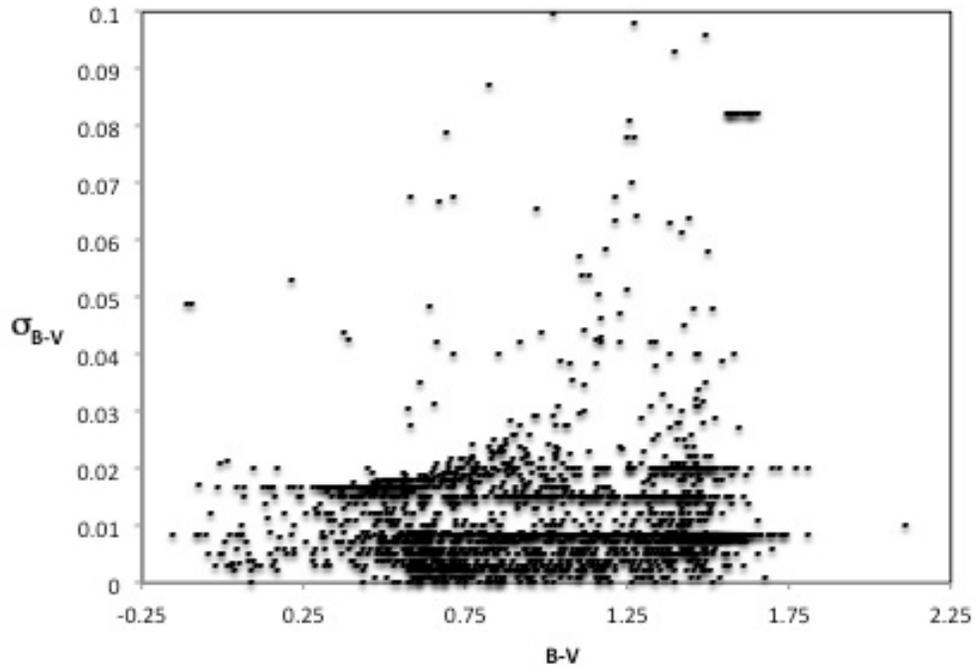

Figure 3. Range of *B-V* colors in ExoCat-1, and associated uncertainties for the various sources of photometry.

To illustrate the relative effects of photometric and astrometric uncertainties, Figures 4 and 5 show the color-magnitude diagram for ExoCat-1, with symbol sizes reflecting uncertainties due to $B-V$ photometry and parallax, respectively. From these figures we see that stars appearing to fall below the main sequence (or above it for red stars) typically have poor measurements; most are likely main sequence stars.

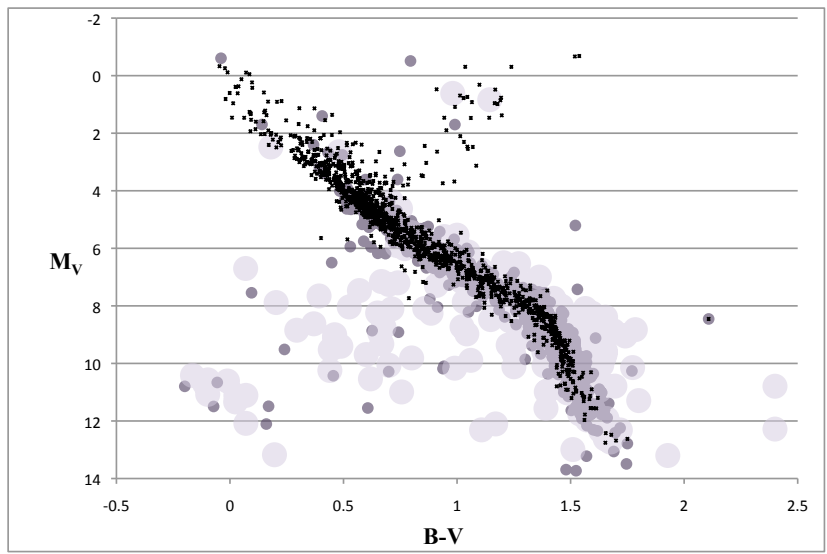

Figure 4.  The effect of photometric uncertainty on the color-magnitude diagram for stars in ExoCat-1.  Symbol sizes reflect uncertainty in B–V of less than 2% (small black dots), 2-5% (medium gray dots), and >5% (large light gray dots).

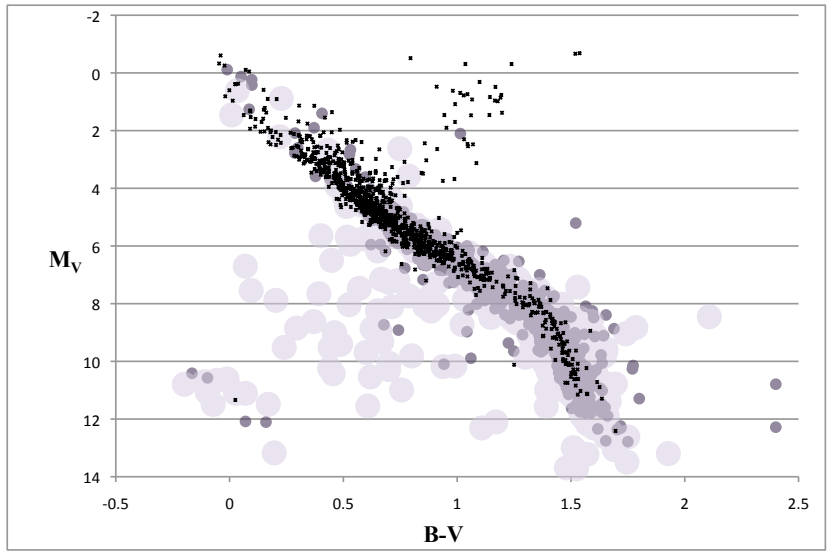

Figure 5.  The effect of fractional parallax uncertainty on the color-magnitude diagram for stars in ExoCat-1.  Symbol sizes as in Figure 4.

### 2.2.3 $K_{(s)}$ Magnitudes from 2MASS and Other Sources

In addition to the optical B and V photometry described above, we sought near-infrared photometry to aid in the estimation of bolometric corrections and luminosity calculations.  The 2MASS All-Sky Catalog of Point Sources is by far the most extensive source of near-infrared photometry available (Cutri et al. 2003), and our preference was to use *2MASS $K_s$* magnitudes whenever possible.  However, we caution that *2MASS* targets are typically saturated at magnitudes brighter than 5-6, which unfortunately includes most of the highest priority targets for direct imaging of habitable exoplanets.  Furthermore, we note that the 2MASS survey utilized a "shortened" bandpass, $K_s$, which differs from the classical Johnson K bandpass in that it excludes wavelengths longward of 2.3 microns, in order to minimize the contribution of thermal background and airglow (Skrutskie et al. 2006).  The *K*-band photometry for bright stars ($K_s$ < 5) in ExoCat-1 comes from a variety of sources, and we attempted to convert all magnitudes to the *2MASS* system.

For a *2MASS $K_s$* magnitude to be included in ExoCat, we required values for the following flags:  (1) quality flag "Qflg" = A (indicating SNR > 10 and measurement uncertainties < 0.10857 mag), (2) read flag "Rflg" = 1 or 2 (indicating that at least one frame in the given bandpass was not overexposed, with the default magnitude derived from either aperture photometry using 51 msec exposure frames for bright, or profile fitting of 1.3 sec frames for fainter stars), and (3) confusion flag "Cflg" = 0 (indicating no apparent problems with bright proximal stars, diffraction spikes or other known artifacts).  This yielded high quality 2MASS $K_s$ data for 1813 ExoCat-1 stars (K_src = "2MASS").  These magnitudes were used directly to calculate $V-K_s$ using the $V$-band data described in the previous section, and the uncertainty in $V-K_s$ color was estimated using the uncertainty quoted in 2MASS, as e_V-K = $\sqrt{(\sigma_V^2 + \sigma_{Ks}^2)}$. Uncertainties in the $K_s$ data from *2MASS* are typically ~0.02 mag.

For stars where *2MASS* data were deemed unreliable, we canvassed the literature for other sources.  $K_s$ data were available for 13 stars from DENIS, which has the advantage of using a bandpass identical to that of 2MASS (Kimeswenger et al. 2004). We quote magnitudes and uncertainties directly for those stars.

Otherwise, we attempted to convert $K$-band data from other systems to that of 2MASS.  Transformations from the SAAO system (which is the classical Johnson system) to 2MASS $K_s$ magnitudes were provided by Koen et al. (2007) and applied to K-band data for 15 stars taken from Koen et al. (2010).  Uncertainties for those stars are quoted at 0.03 mag, which accounts for the internal uncertainty in the measurements as well as in the conversion to 2MASS magnitudes.

We found 41 additional stars in Koorneef (1983) and 23 additional stars in Kidger et al. (2003), who provide transformations to the Koorneef (1983) system.  Carpenter (2003), in turn, provides transformations from Koorneef to the 2MASS $K_s$ system. We estimate 0.05 mag uncertainty in the $K_s$ magnitudes for these stars.

Laney et al. (2012) contains photometry for 7 more ExoCat-1 stars, and while their internal precision is high (0.01 mag), the conversion to the 2MASS system was unclear from the publication.  We opted to include the given K magnitudes with no transformation, with a relatively high assumed uncertainty of 0.05 mag.  Similarly, for 189 stars we found K-band magnitudes from older sources compiled by Gezari et al. (1999), many of which are on the classical Johnson system but generally do not have quoted uncertainties.  For these stars, ExoCat-1 reproduces the $K$ magnitudes directly with no transformation, with a relatively high assumed uncertainty of 0.05 mag.  Finally, we took $K$-band data for 14 stars from the Two Micron Sky Survey (Neugebauer & Leighton 1969).  The given uncertainties in these measurements were quite high (~10%) and we did not attempt to convert these magnitudes to the 2MASS system.

Table 1 lists the references for the above data and the number of measurements taken from each.  This left 244 stars with no K-band data at all, about half of which are quite bright ($V<6$) and potentially of great interest to exoplanet imaging programs.  This situation could be quickly remedied with a dedicated survey incorporating an infrared camera on a small telescope, but until such a survey is carried out, luminosities can be estimated using optical measurements.

Figure 6 shows the spread of $V-K_s$ values and their associated uncertainties for stars in ExoCat-1.  As can be seen from the plot the vast majority of data for redder (and fainter) stars comes from 2MASS and have a natural spread of uncertainties.  For brighter (and bluer) stars, however, the data come from a variety of sources of various quality. We attempted to estimate the uncertainties for these stars somewhat conservatively, given the transformations to *2MASS K$_s$* and each source's quoted uncertainties (where available).

Taken together, Figures 1 through 6 should serve as reminders to the exoplanet imaging community that the results of any concept design study (e.g., Exo-S, WFIRST-AFTA-C) are only as reliable as the input data. In the following subsections, we use this compilation of photometry to estimate stellar luminosities, stellar angular diameters, habitable zone locations, and fractional planet brightnesses of Earth analogs – all of which are important factors in mission planning.

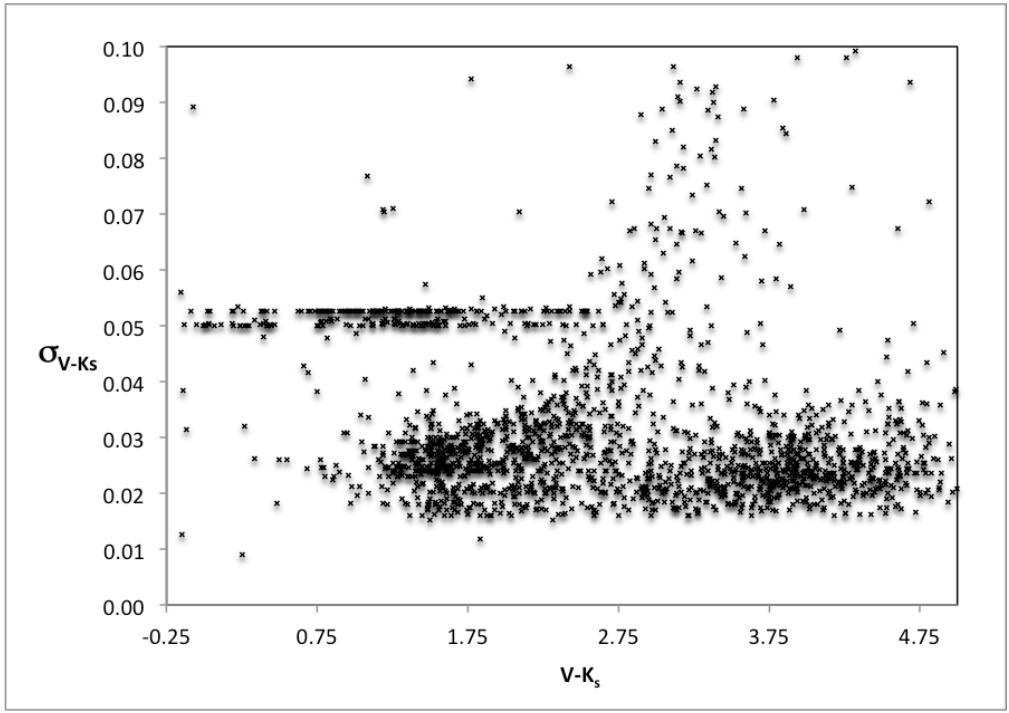

Figure 6. The spread in *V–K$_s$* values and associated uncertainties (in magnitudes) for stars in ExoCat-1. Taken together, Figures 1 through 4 should serve as gentle reminders to the exoplanet imaging community that the results of any concept design study (e.g., Exo-S, WFIRST-AFTA-C) are only as reliable as the input data.

## Table 1
### Sources for *K*-band Data in ExoCat-1

| Reference | # of Stars |
|---|---|
| A'Hearn, M. F., Dwek, E., Tokunaga, A. T. 1984, ApJ, 282, 803 ............................................. | 1 |
| Allen, D. A., Cragg, T. A. 1983, MNRAS 203, 777 ........................................................ | 4 |
| Alonso, A., Arribas, S., & Martinez-Roger, C., A&A Supp, 107, 365 ...................................... | 4 |
| Arribas, S., Martinez-Roger, C., A&A, 215, 305 ........................................................... | 8 |
| Aumann, H. H. & Probst, R. G., ApJ, 368, 264 ............................................................ | 45 |
| Carney, B. W. AJ, 88, 610 ................................................................................ | 3 |
| Carney, B. W. & Aaronson, M. AJ, 84, 867 ............................................................... | 3 |

Carter, B. S. MNRAS, 242, 1 ........................................................................................ 6
2MASS: Cutri et al. 2003 ........................................................................................... 1813
Davidge, T. J. & Simons, D. A. ApJ, 423, 640 ............................................................ 1
Engels, D., Sherwood, W. A., Wamsteker, W., Schultz, G. V. 1981, A&AS, 45, 5 .......... 1
Epchtein, N., Matsuura, O. T., Braz, et al. 1985, A&AS, 61, 203 .............................. 2
Evans II, N. J., Levreault, R. M., Beckwith, S., Skrutskie, M. 1987, ApJ, 320, 364 ........... 1
Groote, D. & Kaufmann, J. P. 1983, A&A Supp, 53, 91 .............................................. 1
Guetter, H. H. 1977, AJ, 82, 598 ................................................................................. 1
Guetter, H. H. 1979, AJ, 84, 1846 .............................................................................. 1
Jameson, R. F. & Akinci, R. 1979, MNRAS, 188, 421 .................................................. 1
Johnson, H. L. 1965, ApJ, 141, 170 ............................................................................ 3
Johnson, H. L. 1965, CoLPL, 3, 73 .............................................................................. 7
Johnson, H. L. 1966, CoLPL, 4, 99 .............................................................................. 52
Johnson, H. L., MacArthur, J. W., & Mitchell, R. I. 1968, ApJ, 152, 465 ...................... 14
Kidger, M. R. & Martín-Luis, F. 2003, AJ, 125, 3311 .................................................. 23
Koen, C., Kilkenny, D., van Wyk, F., & Marang, F. 2010, MNRAS, 403, 1949 .............. 15
Koorneef 1983, A&A, 128, 84 ..................................................................................... 41
Laney, D. D., Joner, M. D., & Pietrzyński, G. 2012, MNRAS, 419, 1637 ...................... 7
Low, F. J., & Johnson, H. L. ApJ, 139, 1130 ................................................................ 4
Mould, J. R. & Hyland, A. R. 1976, ApJ, 208, 399 ...................................................... 1
Neugebauer, G. & Leighton, R. B. 1969, NASA SP-3047 ............................................ 14
Price, S. D. 1968, AJ, 73, 431 ..................................................................................... 2
Selby, M. J., Hepburn, I., Blackwell, D. E. et al. 1988, A&AS, 74, 127, 1988 .............. 8
Storey, J. W. V. 1983, MNRAS, 202, 105 .................................................................... 1
Sylvester, R. J., Skinner, C. J., Barlow, M. J., & Mannings, V. 1996, MNRAS, 279, 915 .......... 1
Veeder, G. J. 1974, AJ, 79, 1056 ................................................................................ 1

## 2.3 Spectral Types, $T_{eff}$, $R_\star$, log(g), $M_\star$, [Fe/H], and Activity Indicators from the Literature

The spectral types ("SpType") given in ExoCat-1 come preferentially from the 40 pc surveys by Gray et al. (2003, 2006) and from the Michigan Catalogue of HD stars for stars south of $\delta = +05°$ (Houk et al. 1975, 1978, 1982, 1988, 1999). If spectral type was not found in those sources, we included the spectral type given in Hipparcos. If Hipparcos did not include a spectral type, we resorted to SIMBAD. For stars without a spectral type in any of those sources, we still provide an estimate of both spectral type (SPECTAG = "O", "B", "A", etc) and a rough luminosity class (CLASS = "GIANT," "MAINSEQ," "SUB," or "WD") based on the star's location on the color-magnitude diagram, as illustrated in Figure 7. This information is provided purely for the user's convenience in color-coding plots of target parameters and at-a-glance information about favorable targets in design studies. In Figure 7 and in ExoCat-1, the SPECTAG values range in $B–V$ following Cox (2000):

B stars: $B–V < 0$; $T_{eff} > 10,000$ K
A stars: $0 \leq B–V < 0.3$; $T_{eff} > 10,000\text{-}7300$ K
F stars: $0.3 \leq B–V < 0.58$; $T_{eff} > 7300\text{-}5940$ K
G stars: $0.58 \leq B–V < 0.8$; $T_{eff} > 5150\text{-}5940$ K
K stars: $0.8 \leq B–V < 1.4$; $T_{eff} > 3840\text{-}5150$ K
M stars: $B–V \geq 1.4$; $T_{eff} > 3840$ K

WD:

$$M_V > 5 \, (B{-}V) + 9 \tag{5}$$

SUB:

$B{-}V < -0.1$: $M_V > 28(B{-}V) + 5.8$

$-0.1 \leq B{-}V < 1.28$: $M_V > 4.8(B{-}V) + 3.5$ $\qquad$ (6)

$1.28 \leq B{-}V$: $M_V > 17(B{-}V) - 12.2$

GIANT:

$$M_V \leq \_10((B{-}V){-}1.4)^2 + 6.5 \tag{7}$$

Effective temperatures, stellar radii, log(g) and masses were either taken from Gray et al. (2003, 2006), Valenti & Fischer (2005), Takeda et al. (2007) or derived from $BVK_s$ data as described in the next Section. Age estimates come either from Takeda et al. (2007) or Valenti & Fischer (2005), rough chromospheric activity indicators (VI= very inactive, I = inactive, A = active, VA = active) come from Gray et al. (2003, 2006), and metallicities ([Fe/H]) come from spectroscopic measurements by Valenti & Fischer (2005). We found that where sources and/or our own derivations overlap, agreement is generally very good.

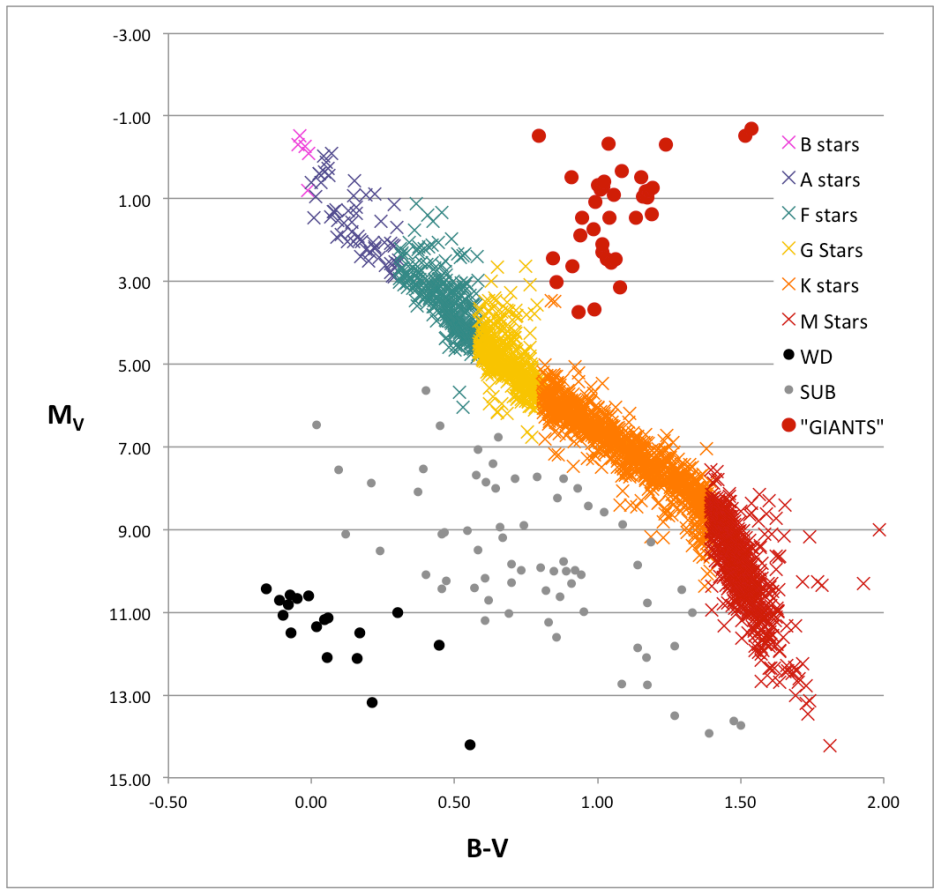

Figure 7. The "SPECTAGS" assigned to stars in ExoCat-1. Note that "SUB" stars are unlikely to be true subdwarfs, but typically have poor quality photometry or parallax data, as discussed in Section 3 and shown in Figures 11 and 12.

## 2.4 Multiple System Data and Known Exoplanets

For known multiple systems, we also have provided magnitudes and separations for the brightest companion within 10", taken from the Washington Double Star catalog (WDS; Mason et al. 2001). For some of these systems, the stability of planet orbits may be called into question, although it is clear that many binary and multiple systems do host exoplanets (e.g., Ragavan et al. 2010). However, background levels from stray light, and required exposure times to detect exoplanets, may be substantially increased for some of these high priority targets (e.g., α Cen B, 70 Oph A, 36 Oph A, η Cas A).

Exoplanet data in ExoCat-1 are limited to the number of known planets in each system. These data come from the NASA Exoplanets Archive of confirmed planets, maintained by the Infrared Processing and Analysis Center at Caltech (http://exoplanetarchive.ipac.caltech.edu/), as of Fall 2013.

## 3. Derived Data: $L_{bol}$, $T_{eff}$, $R_\star$, $M_\star$, EEID, and Planet Brightness

### 3.1 Bolometric Corrections and Luminosity Estimates

The photometry values compiled above were used to estimate bolometric corrections to V-band data ($BC_V$) in order to estimate stellar luminosities, stellar radii and angular size, effective temperature, habitable zone locations, factional planet brightness of Earth-analogs (relative to the star), and apparent magnitudes of Earth-analogs. To accomplish this, we used fits to the $BC_V$ sequence given for O9-M9 dwarfs in Pecaut & Mamajek's (2013) Appendix Table 4, which explicitly includes *2MASS* bands. We estimated $BC_V$ for each star using both $B-V$ and $V-K_s$ (where available).

To find $BC_V$ as a function of $B-V$, the fits we used are as follows:

For $2.13 \geq B-V \geq 1.431$ (M9V – M0V stars):

$$
\begin{aligned}
BC_V = \quad & 338.02061*(B-V)^5 \\
& - 2986.13339*(B-V)^4 \\
& + 10480.01411*(B-V)^3 \\
& - 18266.53141*(B-V)^2 \\
& + 15809.32964*(B-V) \\
& - 5435.53865
\end{aligned}
$$

For $1.431 > B-V \geq 1.1$ (K9V – K4V stars):

$$
\begin{aligned}
BC_V = \quad & 65.931303*(B-V)^4 \\
& - 370.004784*(B-V)^3 \\
& + 767.490707*(B-V)^2 \\
& - 700.601775*(B-V) \\
& + 237.393689
\end{aligned}
$$

For $1.1 > B-V \geq 0.588$ (K3V – G0V stars):

$$
\begin{aligned}
BC_V = \quad & 96.66997*(B-V)^5 \\
& - 403.757274*(B-V)^4 \\
& + 666.340248*(B-V)^3 \\
& - 544.062615*(B-V)^2 \\
& + 219.340924*(B-V) \\
& - 34.955107
\end{aligned} \tag{8}
$$

<u>For 0.588 > $B{-}V \geq 0.16$ (A5V – F9V stars):</u>

$BC_V =$      $71.127951*(B{-}V)^5$

              $- 149.975189*(B{-}V)^4$

              $+ 123.557739*(B{-}V)^3$

              $- 49.974262*(B{-}V)^2$

              $+ 9.724795*(B{-}V)$

              $- 0.721284$

<u>For 0.16 > $B{-}V \geq -0.07$ (B9V – A4V stars):</u>

$BC_V =$      $2.779019*(B{-}V)^4$

              $- 11.93499*(B{-}V)^3$

              $- 5.546819*(B{-}V)^2$

              $+ 2.492754*(B{-}V)$

              $- 0.242511$

<u>For -0.07 > $B{-}V \geq -0.318$ (O9V – B8V stars):</u>

$BC_V =$      $5976.440962*(B{-}V)^5$

              $+ 4547.360571*(B{-}V)^4$

              $+ 1174.286745*(B{-}V)^3$

              $+ 112.335769*(B{-}V)^2$

              $+ 12.796119*(B{-}V)$

              $+ 0.212404$

As shown in Figure 8, the residuals between the above equations and the values in Pecaut & Mamajek's (2013) Table 4 are typically ~1%, except for the O9V – B8V stars ($B{-}V$ < -0.07), where the residuals are as high as 4%.

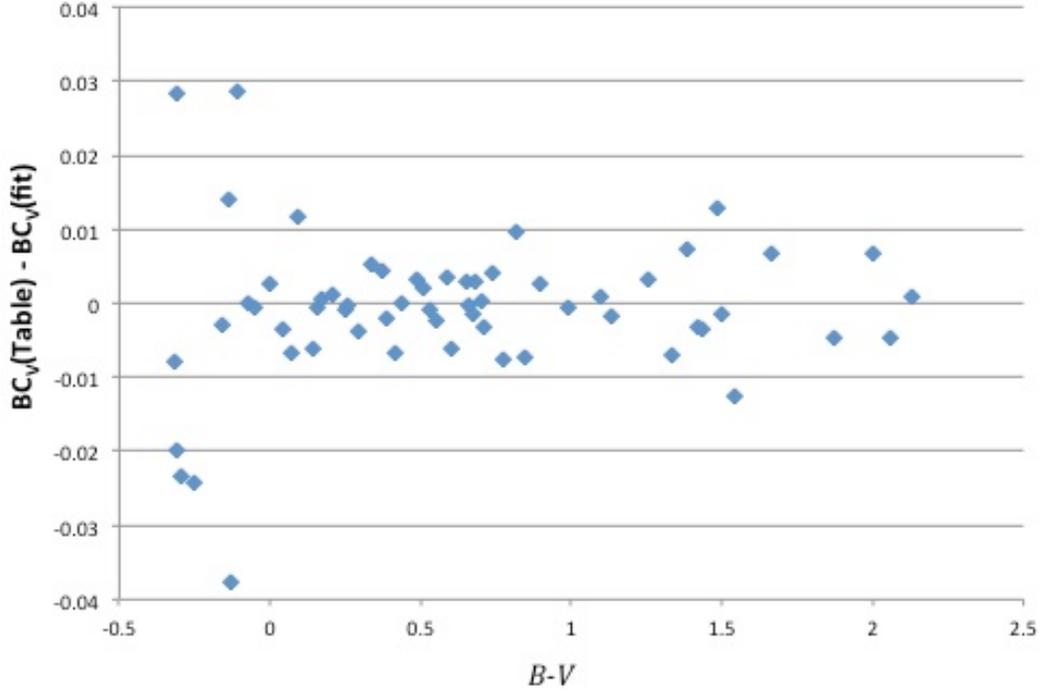

Figure 8. Residuals (in magnitudes) of the fit as a function of $B–V$ to the $BC_V$ values given in Pecaut & Mamajek's (2013) Table 4.

To find $BC_V$ as a function of $V–K_s$, the fits we used are as follows:

For $8.85 ≥ V–K_s ≥ 3.79$ (M9V – M0V stars):

$$
\begin{aligned}
BC_V = \ & 0.00316463*(V–K_s)^6 \\
& - 0.1145736*(V–K_s)^5 \\
& + 1.68936504*(V–K_s)^4 \\
& - 12.96541457*(V–K_s)^3 \\
& + 54.56154672*(V–K_s)^2 \\
& - 120.19018081*(V–K_s) \\
& + 107.98535252
\end{aligned}
$$

For $3.79 > V–K_s ≥ 2.733$ (K9V – K4V stars):

$$
\begin{aligned}
BC_V = \ & -2.374024*(V–K_s)^5 \\
& + 40.046085*(V–K_s)^4 \\
& - 269.601079*(V–K_s)^3 \\
& + 905.169085*(V–K_s)^2 \\
& - 1515.742351*(V–K_s) \\
& + 1012.320734
\end{aligned}
\tag{9}
$$

For  2.733 > $V–K_s$ ≥ 1.421   (K3V – G0V stars):

$BC_V = $  1.0857*$(V–K_s)^5$

- 11.058*$(V–K_s)^4$

+ 44.496*$(V–K_s)^3$

- 88.541*$(V–K_s)^2$

+ 86.912*$(V–K_s)$

- 33.687

For  1.421 > V–Ks ≥ 0.403  (A5V – F9V stars):

$BC_V = $  0.8453*$(V–Ks)^5$

- 4.3756*$(V–K_s)^4$

+ 8.8532*$(V–K_s)^3$

- 8.806*$(V–K_s)^2$

+ 4.2274*$(V–K_s)$

- 0.776

For  0.403 > V–Ks ≥ -0.121   (B9V – A4V stars):

$BC_V = $  -79.113229*$(V–K_s)^5$

+ 83.409453*$(V–K_s)^4$

- 26.407304*$(V–K_s)^3$

- 0.166228*$(V–K_s)^2$

+ 1.660213*$(V–K_s)$

- 0.303442

For  -0.121 > V–Ks ≥ -1  (O9V – B8V stars):

$BC_V = $  -9.560608*$(V–K_s)^6$

- 11.693625*$(V–K_s)^5$

+ 2.547058*$(V–K_s)^4$

+ 3.669271*$(V–K_s)^3$

- 2.648082*$(V–K_s)^2$

+ 1.302543*$(V–K_s)$

- 0.273794

As shown in Figure 9, the residuals between the above equations and the values in Pecaut & Mamajek's (2013) Table 4 are typically ~1% or less, except for the O9V – B8V stars ($V–K_s$ < -0.25), where the residuals are as high as 4%.

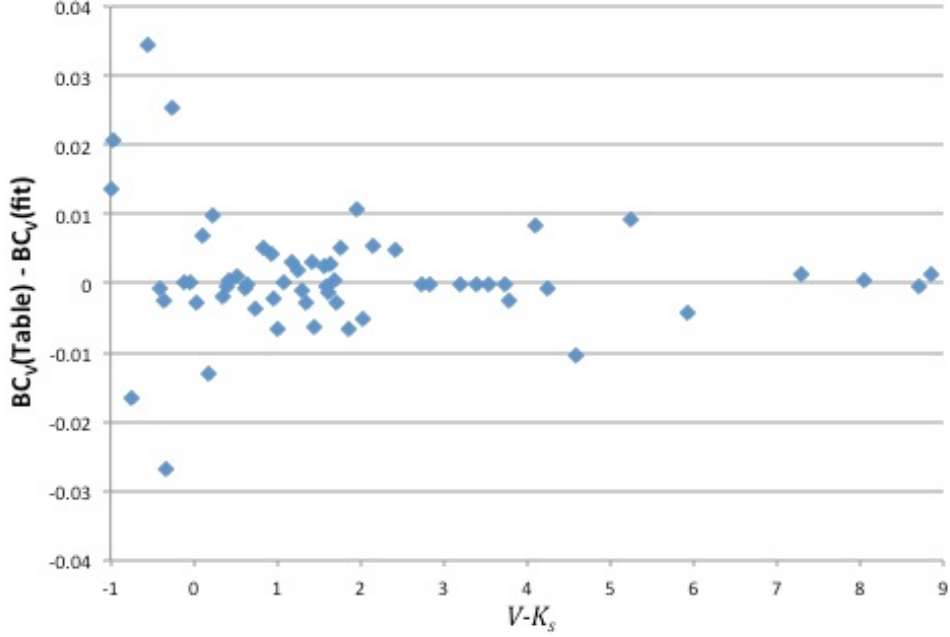

Figure 9. Residuals (in magnitudes) of the fit as a function of $V–K_s$ to the $BC_V$ values given in Pecaut & Mamajek's (2013) Table 4.

To calculate bolometric stellar luminosity from the above bolometric corrections to V-band magnitudes, we used the usual relations:

$$\log (L_\star/L_\odot)_{bol} = -(M_{bol} - M_{bol,\odot})/2.5, \qquad (10)$$

where

$$M_V = V - 5 \log (d(pc)) - 5 = V - 5 \log (1000/\pi(mas)) - 5, \qquad (11)$$

$$M_{bol,\odot} = 4.78, \qquad (12)$$

and

$$M_{bol} = M_V - BC_V. \qquad (13)$$

The bolometric luminosities given in ExoCat-1 generally use $V–K_s$ data to find $BC_V$ from Equations (9)(Lbol_src = "V-K"; 2100 stars). When $V–K_s$ data were not available, $B–V$ data were used with Equations (8)(Lbol_src = "B-V"; 241 stars). For a few stars $V–K_s$ < -1 and highly uncertain, $B–V$ data were also used with Equations (8) (Lbol_src = "B-V"; 3 stars). Where only $V$-band magnitudes were available, no bolometric correction was applied (Lbol_src = "V"; 3 stars).

ExoCat-1 provides the absolute magnitude, $M_V$, and a measure of the uncertainty in $M_V$ (e_Mv) by averaging the effects of adding and subtracting 1-σ from the parallax and V-band measurements. The uncertainty in $M_{bol}$ was then estimated by adding

e_Mv to the maximum residuals for each category of stars in Equations (8) and (9). An estimate of the fractional uncertainty in luminosity is also provided in ExoCat-1 ("e_Lbol") and this value was arrived at by propagating the uncertainty in $M_{bol}$ through Equation (10).

Thus, our knowledge of stellar luminosities is dependent upon parallax, B, V and K photometry, and our knowledge of the bolometric correction sequence. Furthermore, the bolometric corrections used here do not account for differences metallicity or surface gravity. Therefore, in using ExoCat-1 and other sources for target data, it is wise to consider how the uncertainties in luminosity and other parameters might affect the outcome of design studies. Figure 10 shows the range of stellar luminosities and their associated uncertainties in ExoCat-1. Figures 11 and 12 show the HR-diagram and the effects of parallax and photometric uncertainty.

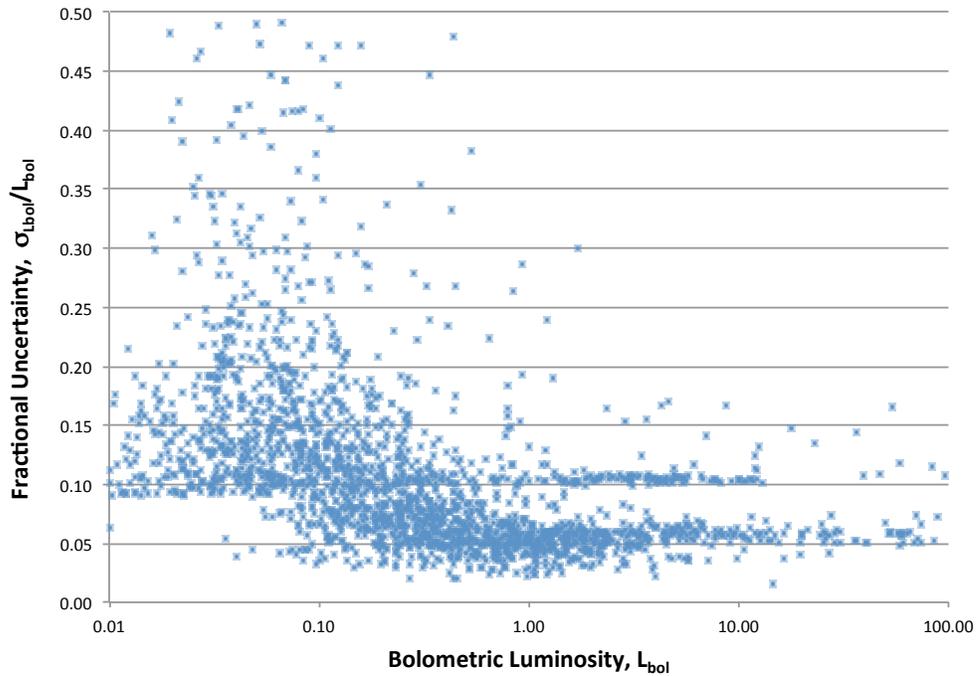

Figure 10. The range of luminosities in ExoCat-1 and associated uncertainty estimates. Uncertainties originate from errors in parallax measurements and BVK photometry, and can be high for both apparently bright stars (due to lack of quality infrared data used here) and faint stars (due to lack of quality optical photometry and lower parallax precision). Stars with close bright companions are also more likely to have higher uncertainties in bolometric luminosity, due to lower precision in the parallax measurement.

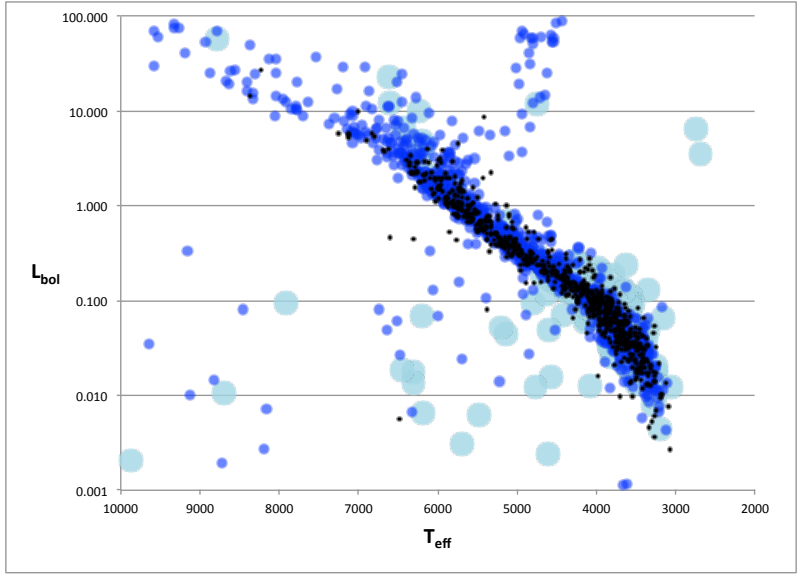

Figure 11. The effect of photometric uncertainty on the HR-diagram for ExoCat-1. Symbol sizes indicate uncertainty in the $BVK_s$ photometry, with stars having all photometric uncertainties less than 2% shown in small black "+" symbols; stars having any of $B$, $V$, or $K_s$ photometric uncertainties between 2-5% shown in small blue circles; and stars having any of $B$, $V$, or $K_s$ photometric uncertainties greater than 5% shown in large light blue circles.

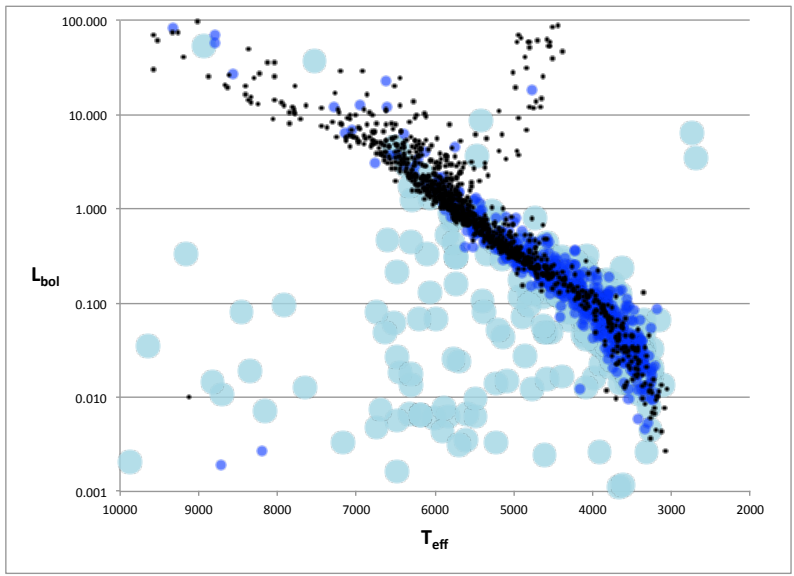

Figure 12. The effect of parallax uncertainty on the HR-diagram for ExoCat-1. Symbol sizes indicate uncertainty in the $BVK_s$ photometry, with stars having all parallax uncertainties less than 2% shown in small black "+" symbols; stars having parallax uncertainties between 2-5% shown in small blue circles; and stars having parallax uncertainties greater than 5% shown in large light blue circles.

## 3.2 Effective Temperatures, Radii, and Masses

The effective temperatures given in ExoCat-1 and used in Figures 11 and 12 come either from the literature (as described in Section 2.3), or were derived using fits from Table 4 of Pecaut & Mamajek (2013) either as a function of $V{-}K_s$ or as a function of $B{-}V$ (for hotter stars with $V{-}K_s < 0.4$, or where $V{-}K_s$ data were lacking). The equations we used in these fits are as follows:

Underline: For $V{-}K_s \geq 0.4$:

$$
\begin{aligned}
T_{eff} = \ & 0.1503*(V{-}K_s)^6 \\
& - 4.8644*(V{-}K_s)^5 \\
& + 62.204*(V{-}K_s)^4 \\
& - 409.26*(V{-}K_s)^3 \\
& + 1570.3*(V{-}K_s)^2 \\
& - 4071.6*(V{-}K_s) \\
& + 9510.1
\end{aligned}
\tag{14}
$$

Residuals between the above equation and Table 4 of Pecaut & Mamajek (2013) are less than 50K.

For $B{-}V < 0.29$:

$$
\begin{aligned}
T_{eff} = \ & 2376073.8887*(B{-}V)^6 \\
& - 14456.8145*(B{-}V)^5 \\
& - 260156.155*(B{-}V)^4 \\
& - 228921.4859*(B{-}V)^3 \\
& + 106909.3718*(B{-}V)^2 \\
& - 18098.381*(B{-}V) \\
& + 9433.1809
\end{aligned}
\tag{15}
$$

For $B{-}V > 0.29$:

$$
\begin{aligned}
T_{eff} = \ & -224.47*(B{-}V)^4 \\
& - 660.86*(B{-}V)^3 \\
& + 3930.2*(B{-}V)^2 \\
& - 7245.3*(B{-}V)
\end{aligned}
\tag{16}
$$

+ 9009

Residuals between the above equations and Table 4 of Pecaut & Mamajek (2013) are less than 100K for stars cooler than ~8500 K, up to a few hundred degrees for the hottest stars in ExoCat-1 (~15,000 K).

Finally, we found stellar radii according to:

$(R_\star/R_\odot) = \sqrt{(L_\star/L_\odot)_{bol}} \; (T_{eff}/T_{eff_\odot})^2$, where $T_{eff_\odot}$ = 5770 K,           (17)

and approximate stellar masses:

$(M_\star/M_\odot) \approx (L_\star/L_\odot)^{1/3.9}$,           (18)

which is a reasonable approximation for stars between 0.5 – 20 $M_\odot$ (Salaris & Cassisi 2005).

### 3.3 Quantities for Earth Analogs: EEID, FPB, and V-magnitudes

To estimate the location of the habitable zone, we found the separation where an Earth twin would receive the same amount of energy from its parent star as the Earth does from the Sun. ExoCat-1 includes this "Earth-Equivalent Insolation Distance" (EEID) for each star, calculated as follows:

$EEID \; (AU) = \sqrt{L_{bol}}$,

and

$EEID \; (mas) = 1000 * EEID \; (AU) / d(pc).$           (19)

The fractional planet brightness (FPB) of that planet relative to its star is calculated in ExoCat-1 following Turnbull et al. (2012):

$FPB = (F_p/F_\star)_{EEID} = 1.155 \times 10^{-10}/L_{bol}$,           (20)

for a Bond albedo of 0.3, at quadrature (90° phase angle). The V-band magnitude of the Earth twin is then:

$V_\oplus = -2.5 \log (FPB) + V$,           (21)

where $V$ is the $V$-band magnitude of the star itself.

In Figure 13, we show the ranges of fractional planet brightness and separations for Earth twins orbiting main sequence stars in ExoCat-1, color-coded by the planets' $V$-band magnitudes, which span from $V_\oplus$ = 25th – 35th magnitude.

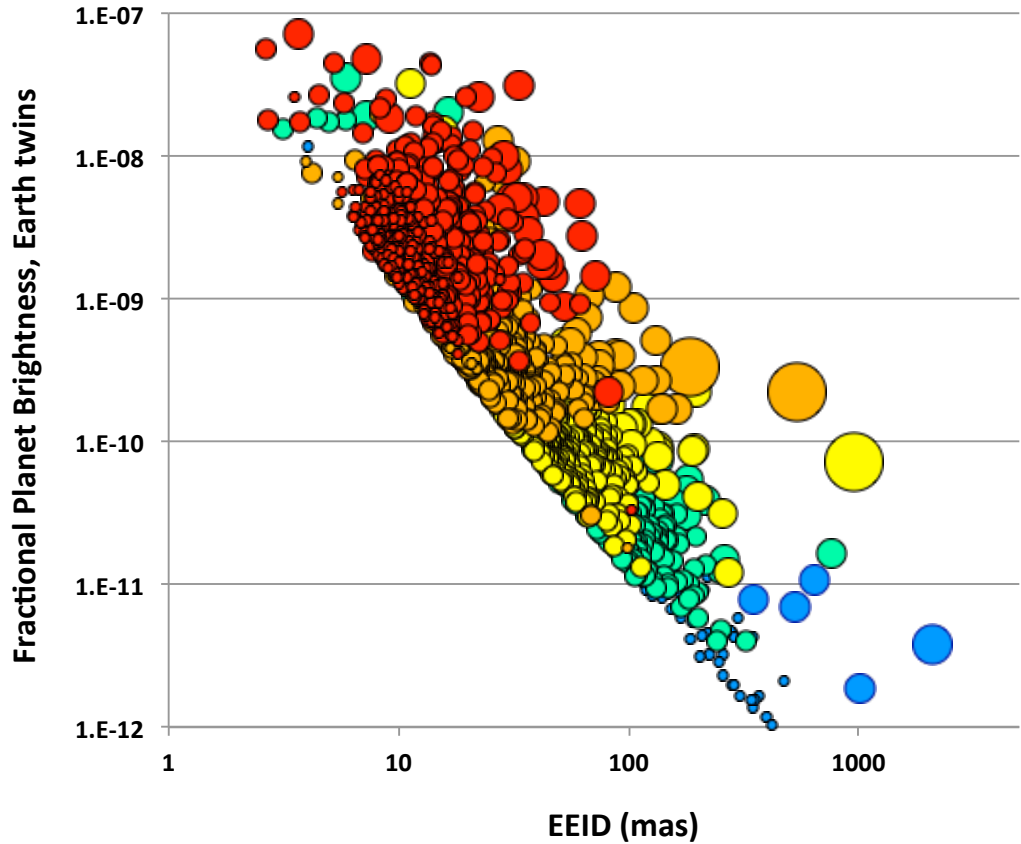

Figure 13. Fractional Planet Brightness for Earth Twins at the EEIDs of the main sequence stars in ExoCat-1. Symbols are color-coded by spectral type (O/A = blue, F = green, G = yellow, K = orange, M = red) and sized by apparent *V*-band magnitude of the hypothetical Earth twin (large circles: *V* < 27.5 mag; medium circles: 27.5 < V < 30 mag; small circles: 30 < *V* < 32.5 mag; tiny circles: 32.5 < *V* < 35 mag).

## Section 4.  Catalog Maintenance and Future Directions

A short explanation of columns for ExoCat-1 is given in Appendix A.  The full catalog can be accessed online through the Exoplanets Exploration Program (ExEP) website (http://nexsci.caltech.edu/missions/EXEP/EXEPstarlist.html).  Additions and corrections to the catalog will be made by the author for the foreseeable future, and high-priority improvements to ExoCat include:

- deriving bolometric corrections from accurate spectral types (preferentially from Gray et al. 2003, 2006),

- adding later type stars with high quality ground-based parallax data,

- including separate entries for all known companions in each system,

- adding flags on suspected multiple systems, including unresolved systems,

- including companion data for all stars in the WDS catalog,
- adding Johnson U, I, J, and H-band data wherever available, especially for high priority exoplanet imaging and transit targets, and
- adding a flag for targets with known circumstellar debris.

**Acknowledgements.** The author thanks B. Skiff (Lowell Obs.) for many helpful discussions in wading through the endless vagueries of the astronomical literature; R. Brown (STScI) and R. Trabert (JPL) for suggestions, questions, and feedback as ExoCat "power users;" and most importantly, W. Traub and the Exoplanet Exploration Program office (JPL) for seeing both the value and inherent difficulties in creating such a compilation. She also thanks E. Mamajek (U. Rochester) for discussions regarding infrared photometry and its use in estimating bolometric luminosities. This work was supported primarily by the NASA Exoplanet Exploration Program and secondarily by the NASA Astrobiology Institute via an NAI DDF award through the University of Arizona's "Follow the Elements" NAI Team (PI: A. Anbar). This research has made use of the Washington Double Star Catalog maintained at the U.S. Naval Observatory.

**REFERENCES (in addition to those listed in Table 1)**

Bakos, G. A., Sahu, K. C., & Nemeth, P. 2002, ApJS, 141, 187

Bessell, M. 2000, "The *Hipparcos* and *Tycho* Photometric System Passbands," PASP, 112, 961

Bessell, M. 2005, "Standard Photometric Systems," ARAA, 43, 293

Brown, R. 2006, "Expectations for the Early TPF-C Mission," in Direct Imaging of Exoplanets: Science & Techniques. Proceedings of the IAU Colloquium #200, Edited by C. Aime and F. Vakili. Cambridge, UK: Cambridge University Press, 2006., pp.119-128

Brown, R. 2015, "Science Parametrics for Missions to Search for Earth-like Exoplanets by Direct Imaging," ApJ, 799, 87

Burrows, A. S. 2014, "Spectra as windows into exoplanet atmospheres," PNAS, 111, 12601

Cannon, A.J., Pickering, E.C.: Annals of Harvard College Observatory 97, 1 (1922)

Dommanget, J., & Nys, O. 2000,"Erratum: The visual double stars observed by the Hipparcos satellite," AA, 364, 927

ESA. 1997a, "The *Hipparcos* and Tycho Catalogues," (ESA SP-1200; Noordwijk: ESA)

Gezari, D. Y, Pitts, P.S., & Schmitz, M. 1999, "Catalog of Infrared Observations, Edition 5," VizieR On-line Data Catalog: II/225

Gliese, W., & Jahreiss, H. 1995, VizieR On-line Data Catalog: V/70A. Originally published in: Astron. Rechen-Institut, 5070, 0

Gray R. O., Corbally C. J., Garrison R. F., Mcfadden M. T., Robinson P. E., 2003, "Contributions to the NStars Program: Spectroscopy and Physical Properties of the Solar-type Stars within 40 parsecs of the Sun," AJ, 126, 2048

Gray R. O., Corbally C. J., Garrison R. F., McFadden M. T., Bubar E. J., McGahee C. E., O'Donoghue A. A., Knox E. R., 2006, "Contributions to the Nearby Stars (NStars) Project: Spectroscopy of Stars Earlier than M0 within 40 pc-The Southern Sample," AJ, 132, 161

Henry, T. J.; Jao, W.; Subasavage, J. P.; Beaulieu, T. D.; Ianna, P. A. et al. 2006, AJ, 132, 2360

Høg, E., Fabricius, C., Makarov, V.V., Urban, S., Corbin, T., Wycoff, G., Bastian, U., Schwekendiek, P., Wicenec, A. 2000, "The Tycho-2 Catalogue of the 2.5 Million Brightest Stars," A&A, 355, L27

Houk N., 1978, Michigan Catalogue of Two-dimensional Spectral Types for the HD Stars Vol. 2, Declinations −53◦.0 to −40◦.0. Department of Astronomy, University of Michigan, Ann Arbor, MI, USA

Houk N., 1982, Michigan Catalogue of Two-dimensional Spectral Types for the HD Stars Vol. 3, Declinations −40◦.0 to −26◦.0. Department of Astronomy, University of Michigan, Ann Arbor, MI, USA

Houk N. & Cowley A. P., 1975, Michigan Catalogue of Two-dimensional Spectral Types for the HD stars Vol. 1, Declinations −90◦.0 to −53◦.0. Department of Astronomy, University of Michigan, Ann Arbor, MI, USA

Houk N. & Smith-Moore M., 1988, Michigan Catalogue of Two-dimensional Spectral Types for the HD Stars Vol. 4, Declinations −26◦.0 to −12◦.0. Department of Astronomy, University of Michigan, Ann Arbor, MI, USA

Houk N. & Swift C., 1999, Michigan Catalogue of Two-dimensional Spectral Types for the HD Stars Vol. 4, Declinations −12◦. 0 to +05◦. 0. Department of Astronomy, University of Michigan, Ann Arbor, MI, USA

Howard, A. W. 2013, "Observed Properties of Exoplanets," Science, 340, 572

Jordi, C. 2015, "Gaia is now a reality," Highlights of Spanish Astrophysics VIII, Proceedings of the XI Scientific Meeting of the Spanish Astronomical Society held on September 8-12, 2014, in Teruel, Spain, ISBN 978-84-606-8760-3. A. J. Cenarro, F. Figueras, C. Hernández-Monteagudo, J. Trujillo Bueno, and L. Valdivielso (eds.), pp. 390-401

Koen, C., Marang, F., Kilkenny, D., & Jacobs, C. 2007, "Improved SAAO-2MASS photometry transformations," MNRAS, 380, 1433

Laughlin, G. & Lissauer J. L. 2015, Review chapter to appear in Treatise on Geophysics, 2nd Edition, arXiv:1501.05685

Léger, A., Defrere, D., Malbet, F., Labadie, L., Absil, O. 2015, "Impact of {eta}earth on the capabilities of affordable space missions to detect biosignatures on extrasolar planets," ApJ, accepted

Lindegren, L.; Mignard, F.; Söderhjelm, S.; Badiali, M.; Bernstein, H.-H. et al. 1997, "Double star data in the HIPPARCOS Catalogue," A&A, 323, 53

van Leeuwan, F. 2007, "*Hipparcos*, the new Reduction of the Raw Data," Astrophysics and Space Science Library, Vol. 350, pp. 113-115. Berin: Springer, ISBN# 978-1-40206342-8

Luyten, W. J. 1976, "A Catalogue of 1849 Stars with Proper Motions Greater than 0.5″ Annually (LHS)", Univ. Minnesota

Luyten, W. J., 1979, "New Luyten catalogue of stars with proper motions larger than two tenths of an arcsecond; and first supplement; NLTT" (Minneapolis (1979)), Vizier CDS Catalog I/87

Makarov, V. V. 1997, Proceedings of the ESA Symposium `Hipparcos - Venice '97', 13-16 May, Venice, Italy, ESA SP-402 (July 1997), pp. 541-544

Mamajek, E. E., Meyer, M. R., & Liebert, J. 2006, "Erratum: Post-T Tauri Stars in the Nearest OB Association," AJ, 131, 2360

Mason, B. D., Wycoff, G. L., Hartkopf, W. I., Douglass, G. G., & Worley, C. E. 2001, "The 2001 US Naval Observatory double star CD-ROM. I. The Washington double star catalog," AJ, 122, 3466

NRC 2010. "New Worlds, New Horizons in Astronomy and Astrophysics," ISBN# 978-0-309-15802-2, http://www.nap.edu/catalog/12951/new-worlds-new-horizons-in-astronomy-and-astrophysics

Pecault, M. J. & Mamajek, E. 2013, "Intrinsic Colors, Temperatures, and Bolometric Corrections of Pre-main-sequence Stars," ApJ, 208, 9

Platais, I., Poubaix, D., Jorissen, A., et al. 2003, "Hipparcos red stars in the $H_pV_{T2}$ and $VI_C$ systems," A&A, 397, 997

RECONS (Research Consortium On Nearby Stars). 2012, "The One Hundred Nearest Star Systems," http://www.recons.org/TOP100.posted.htm

Salaris, M. & Cassisi, S. 2005. "Evolution of stars and stellar populations," John Wiley & Sons. pp. 138 – 140.  ISBN 0-470-09220-3

Seager, S. and the Exo-S Science and Technology Definition Team, 2015. "Exo-S Starshade Probe-Class Exoplanet Direct Imaging Mission Cencept Final Report," https://exep.jpl.nasa.gov/stdt/Exo-S_Starshade_Probe_Class_Final_Report_150312_URS250118.pdf

Spergel, D., Gehrels, N., the WFIRST-AFTA Science Definition Team, and the WFIRST Study Office, 2015.  "Wide-Field InfraRed Survey Telescope-Astrophysics Focused Telescope Assets WFIRST-AFTA 2015 Report," http://wfirst.gsfc.nasa.gov/science/sdt_public/WFIRST-AFTA_SDT_Report_150310_Final.pdf

Stapelfelt, K. and the Exo-C Science Definition Team, 2015. "Exo-C: Imaging Nearby Worlds," https://exep.jpl.nasa.gov/stdt/Exo-C_Final_Report_for_Unlimited_Release_150323.pdf

Takeda, G., Ford, E. B., Sills, A., Rasio, F. A., Fischer, D. A., & Valenti, J. A. 2007, "Structure and evolution of nearby stars with planets. II. Physical properties of ~1000 cool stars from the SPOCS catalog," ApJS, 168, 297

Turon, C. Crézé, M.; Egret, D.; Gómez, A. E.; Grenon, M.; Jahreiß, H.; et al. (1992). *Hipparcos* Input Catalogue, ESA SP-1136 (7 volumes). European Space Agency.

Valenti, J. A., & Fischer, D. A. 2005, "Spectroscopic Properties of Cool Stars (SPOCS). I. 1040 F, G, and K Dwarfs from Keck, Lick, and AAT Planet Search Programs," ApJS, 159, 141

## Appendix – Explanation of ExoCat Columns

HIP     HD      GL/GJ  GL/LTT        COMMON

Common catalog identifiers and names.

WDS

Washington Double Star survey identifier.  This star may have multiple physical and/or optical companions, as listed in the WDS catalog.

sep(") dM(mag)

Given for bright (V<7) targets only.  Separation and delta-magnitude of nearest bright companion, taken from WDS.  Companion must have dM<5 (1%) and be within 30 arcseconds of the target in order to be given here.  Other physical or optical companions may be listed in WDS.

NPLANETS

Number of currently known planetary companions as found in various sources.

RAhms        DEdms         RA(ICRS)      DE(ICRS)      pmRA pmDE Glon   Glat

HipCat Coordinates, proper motions, galactic latitude and longitude.

d(pc)         ePARX/PARX

Distance and fractional parallax uncertainty.  Stars with eparx/parx > 10% have been colored RED in the spreadsheet as a warning that all data are suspect.  Some of these stars occasionally show up as "good" targets.

V        e_V      V_src  B–V     e_B–V  B–V_src

Photometry as transformed to Johnson from *Hipparcos* or ground based data, whichever was higher quality.

Mv      e_Mv    CombMag

Absolute magnitude and uncertainty in magnitudes, flag (*) on entries where more than one star contributes to the measurement.

V-K      e_V-K   K_src

V-K color, uncertainty in magnitudes, and source of K-band photometry.  2MASS is not generally of good quality for bright stars, and thus many other sources had to be compiled.  Wherever possible, those sources were transformed to the 2MASS system, and this was possible for recent measurements.  For many of the brightest stars, however, the best photometry available is decades old.  Attempts to transform those measurements to a single system would likely introduce more uncertainties than are differences between the systems.  Those measurements are quoted as-is.

Lbol_src       Lbol   e_Lbol SpType       CLASS SPECTAG

Bolometric luminosity, using bolometric corrections derived either from V-K (cooler stars) or B–V data (hotter stars).  Spectral Types taken preferentially from Gray+03/06, Houk catalogs, or SIMBAD.  CLASS and SPECTAG are convenient for color-coding plots:  CLASS is either WD, SUB, MAINSEQ, or GIANT indicating rough placement on the HR diagram, and SPECTAG is O, B, A, F, G, K, M, GIANT, SUB, or WD.

EEID(AU)       EEID(mas)

Earth-equivalent Insolation Distance is the distance where an Earth-sized planet would receive the same incoming stellar radiation as does the Earth.  Uses the star's bolometric luminosity.  This is a fast proxy for hab zone location without complications regarding inner and outer HZ edges.

FPB-Earth

Fractional planet brightness of an Earth-sized and albedo planet at the EEID at quadrature, relative to the star in V-band.

V-Earth(mag)

V-band magnitude of Earth-sized and albedo planet at the EEID at quadrature.

----STELLAR PARAMETERS------

Teff    Teff_src       R*(Rsun)     R_src  M*(Msun)    M_src

Effective Temperature in Kelvin, stellar radius and mass in solar units.  These quantities were preferentially taken from fits to high quality spectral data by Takeda+07, Valenti&Fischer05, Gray+03/06, or derived by Turnbull.  Agreement across sources and derivations is very good.

log(g)  log(g)_src    Age(Gyr)      Age_src       AC-Gray       [Fe/H]-VF05

Gravity, very approximate age in Gyr, chromosperic Activity Class from Gray+03/06, and metallicity from VF05.  AC-Gray flag refers to chromospheric Activity Class as follows:

VI =    very inactive

I =      inactive

A =      active

VA =     very active

This is sometimes used as a proxy for age.